\documentclass[numerical,10pt,superscriptaddress]{revtex4-2}
\usepackage{graphicx}
\usepackage{amsmath,amssymb}
\usepackage{times,mathptmx}
\usepackage{units}
\usepackage{epsfig}
\usepackage{booktabs}
\usepackage{multirow}

\usepackage[english]{babel}
\usepackage{color}
\usepackage[symbol]{footmisc}

\newcommand{\VUB}{Applied Physics Research Group, Vrije Universiteit Brussel, Pleinlaan 2, 1050 Brussels, Belgium}

\begin{document}

\date{\today}
\title{Noise-injected analog Ising machines enable ultrafast statistical sampling and machine learning}

\author{Fabian B{\"o}hm}\email[Corresponding authors: ]{Fabian B\"ohm (fabian.bohm@vub.be), Guy Van der Sande (Guy.Van.der.Sande@vub.be)}\affiliation{\VUB}
\author{Diego Alonso-Urquijo}\affiliation{\VUB}
\author{Guy Verschaffelt}\affiliation{\VUB}
\author{Guy Van der Sande}\email[Corresponding authors: ]{Fabian B\"ohm (fabian.bohm@vub.be), Guy Van der Sande (Guy.Van.der.Sande@vub.be)} \affiliation{\VUB}

\begin{abstract}

Ising machines are a promising non-von-Neumann computational concept for neural network training and combinatorial optimization. However, while various neural networks can be implemented with Ising machines, their inability to perform fast statistical sampling makes them inefficient for training neural networks compared to digital computers. Here, we introduce a universal concept to achieve ultrafast statistical sampling with analog Ising machines by injecting noise. With an opto-electronic Ising machine, we experimentally demonstrate that this can be used for accurate sampling of Boltzmann distributions and for unsupervised training of neural networks, with equal accuracy as software-based training. Through simulations, we find that Ising machines can perform statistical sampling orders-of-magnitudes faster than software-based methods. This enables the use of Ising machines beyond combinatorial optimization and makes them into efficient tools for machine learning and other applications.

\end{abstract}
\maketitle
\newpage

\section*{Introduction}

Machine learning with neural networks has led to a revolution in our capabilities to process and analyze large sets of complex data and has become essential, e.g., for machine vision, traffic and stock market prediction or autonomous vehicle control. On the flipside of this revolution is the fact that training neural networks is a computationally expensive task that has to be performed on resource-intensive high-performance computing hardware. This is starting to raise serious concerns about economical and ecological sustainability \cite{XU18,STR19}, which has instigated an intensive search for alternative computing systems, such as quantum annealers or hybrid analog-digital computing concepts \cite{JOH11,CAI20,PRA20,PIE20}, that can perform training of neural networks significantly faster and more efficiently than current generations of digital computers \cite{ZIE20,SAN17}. Among this drive for more efficient computing concepts, analog Ising machines have emerged as a promising solution \cite{YAM17a,VAD20}. They work based on the insight that various difficult computational problems can be mapped to a simple spin system, the so called Ising model \cite{LUC14}, and implemented with artificial spin networks based on analog physical systems \cite{INA16a,BER17,CHO19,BOE19}. Through their natural tendency to evolve to their lowest energy configuration, analog Ising machines are able to find solutions, which allows to forgo many of the limitations of von-Neumann-based computing platforms and has resulted in accelerated computation of difficult combinatorial optimization problems \cite{CAI20,INA16a,HON21,ALB18}. However, while various neural network architectures can be mapped to analog Ising machines \cite{ACK85,BEN16,ULA19}, training these models is currently very inefficient and much slower compared to training on a digital computer. This is because training these neural networks requires Boltzmann sampling, which is the estimation of the neuron activation probabilities in a thermal equilibrium state. As analog Ising machines are known to naturally implement spin systems at very low temperatures, they cannot reach a thermal equilibrium at arbitrary temperatures and are thus unable to perform Boltzmann sampling \cite{BOE18}. While alternative sampling methods based on trapping in local energy minima have been proposed \cite{BEN16,ULA19,SAK16,LIU18}, these methods possess serious drawbacks due to their inaccuracy \cite{BOE18}, complex temperature control schemes \cite{BEN16} and their large performance overhead \cite{LIU18}, which makes Ising machine-based Boltzmann sampling slow and uncompetitive against software-based methods.

To enable efficient Boltzmann sampling with analog Ising machines and leverage their inherent speed for accelerating machine learning, we propose a universal way of performing statistical sampling with analog Ising machines, where noise from an analog noise source is injected to drive the Ising machine into a thermal equilibrium state. Based on a time-multiplexed opto-electronic Ising machine \cite{BOE19}, we experimentally demonstrate how this can be used to efficiently generate samples and to approximate statistical distribution functions with high accuracy. We also demonstrate the application of such noise-induced sampling to the unsupervised training of neural networks and experimentally show that Ising machines are able to achieve accuracy equal to software-based training methods in an image recognition task. Moreover, we numerically estimate the sampling rate of spatially-multiplexed analog Ising machines and find that, with off-the-shelf components, a single analog Ising machine is able to attain GSamples/s sampling rates even for complex large-scale problems. Compared to existing Ising machine-based and software-based sampling methods, this presents a speedup of the sampling speed by several orders-of-magnitude. As Boltzmann sampling is a computationally highly expensive task during neural network training, noise-induced sampling presents a way to leverage the inherent speed of analog Ising machines for more efficient training. Furthermore, as Boltzmann sampling is ubiquitous in various applications, such as drug research or finance \cite{SAK16,WIL07}, it opens up the use of analog Ising machines in fields beyond combinatorial optimization.

\section*{Results}

\subsection*{Noise-induced Boltzmann sampling with analog Ising machines}

\begin{figure}[htbp]
	\centering
		\includegraphics[width=.99\textwidth]{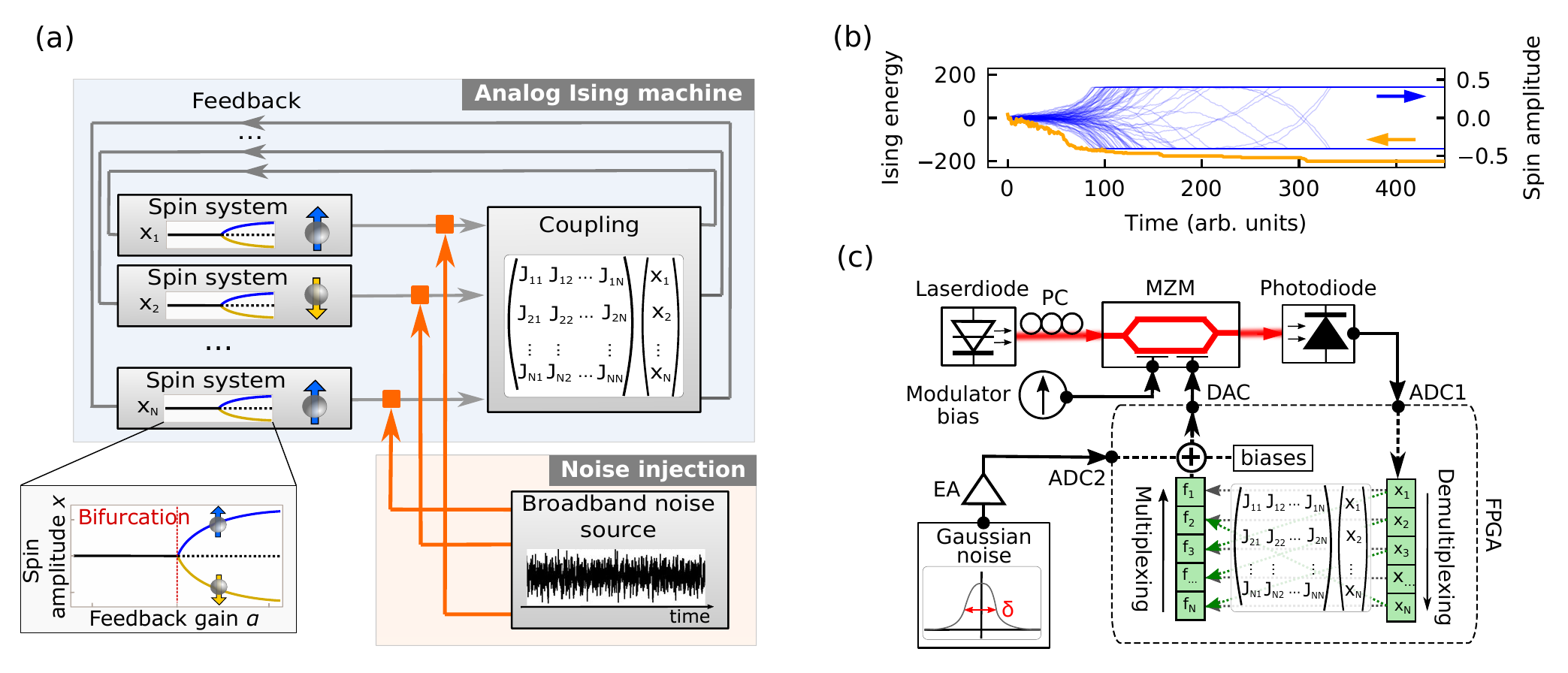}
	\caption{\textbf{Schematic of noise-induced sampling with spatially-multiplexed and time-multiplexed analog Ising machines} (a) Schematic of a spatially multiplexed analog Ising machine for noise-induced sampling. The system consists of a set of $N$ bistable nonlinear systems that represent $N$ spin states and are mutually coupled. Inset: Bifurcation diagram of the spin amplitude as a function of the feedback gain for a single bistable system. Below the bifurcation point at $\alpha=1$ (red dashed line), only the trivial solution exists (solid black line), while above the bifurcation point, the trivial solution becomes unstable (black dotted line) and two new bistable fixed points arise (orange and blue line). (b) Exemplary time evolution of the Ising energy (orange) and the spin amplitudes (blue) while solving a Maxcut optimization problem with $N=100$ spins. (c) Experimental setup of the time-multiplexed opto-electronic Ising machine, where Gaussian white noise with a standard deviation of $\delta$ is injected. PC polarization controller, ADC analog–digital converter, DAC digital–analog converter, MZM Mach-Zehnder modulator, EA electronic amplifier, FPGA field programmable gated array.}
	\label{fig1}
\end{figure}

Boltzmann sampling is the task of approximating the Boltzmann distribution function for an ensemble of $N$ systems

\begin{equation}
	P(\{\sigma\}_i)=\frac{e^{-E(\{\sigma\}_i)/T}}{\sum_q e^{-E(\{\sigma\}_q)/T}} \ \ ,   \label{eq1}
\end{equation}

where $P(\{\sigma\}_i)$ is the probability of measuring a state $\{\sigma\}_i=\{\sigma_1,\sigma_2,...,\sigma_N\}$ with the corresponding energy $E(\{\sigma\}_i)$ at a given temperature $T$. Here, we assume $T$ and $E$ to be dimensionless. For the Boltzmann distribution of an Ising spin system, the energy $E(\{\sigma\}_i)$ is given by the Ising Hamiltonian

\begin{equation}
	E_\mathrm{Ising}(\{\sigma\}_i)=-\frac{1}{2} \sum_{mn}J_{mn}\sigma_m\sigma_n - \sum_m b_m \sigma_m \ \ .\label{eq2}
\end{equation}

Such an Ising model describes a network of binary spins $\sigma_m$ that can point either spin up ($\sigma_m=1$) or spin down ($\sigma_m=-1$). The spins are mutually coupled through the coupling matrix $J_{mn}$ and are subjected to biases $b_m$. Analog Ising machines are physical systems that implement the Ising model, such that their energy is equivalent to equation \eqref{eq2}. Fig.\ref{fig1}a shows a generalized schematic of a spatially-multiplexed analog Ising machine. Such an Ising machine is a feedback system that consists of $N$ parallel nonlinear systems with analog amplitudes $x_m$, which are used to represent $N$ spin states \cite{BOE21}. Contrary to other types of Ising machines with digital spin states \cite{CAI20,PRA20,PIE20}, the binary spins are encoded in a physical system with a continuous variable. To map the binary Ising spins $\sigma_m$ to the analog variables $x_m$, the individual nonlinear systems possess a symmetrical bistability, which is induced by a pitchfork-type bifurcation. A typical bifurcation diagram of the amplitude $x_m$ as a function of the feedback gain $\alpha$ is shown in the inset in Fig.\ref{fig1}a. Spin states are then mapped to these bistable states by taking the sign of the amplitude $\sigma_m=\mathrm{sign}(x_m)$. To use such spin systems for computation, analog Ising machines possess a coupling system, which couples the different nonlinear system according to $J_{mn}$ and thereby implements the Ising Hamiltonian \cite{LEL17,BOE21}. Fig.\ref{fig1}b shows an exemplary time evolution of the Ising energy and the spin amplitudes $x_m$ when emulating a 2D spin lattice with antiferromagnetic ($J_{mn}=-1$) nearest neighbor coupling for $N=100$ spins with an analog Ising machine. When initialized in a random state, the coupling leads to a reordering of the spins which, in turn, minimizes the Ising energy. This reordering continues until the system converges to a stable configuration corresponding to a minimum of the Ising Hamiltonian. In general, analog Ising machines are known to implement the Ising model at very low temperatures ${T}$, which ensures that the final configuration is at or very close to the global energy minimum of equation \eqref{eq2} \cite{BOE18}. While this makes them inherently suitable for solving optimization problems, it also presents a significant drawback for Boltzmann sampling, which requires the ability to emulate the Ising model at any arbitrary temperature. In order to emulate thermal equilibrium dynamics with analog Ising machines, we propose to inject broadband noise into the system in Fig.\ref{fig1}a. Contrary to realistic noise conditions in analog Ising machines, where the standard deviation of the noise distribution $\delta$ is significantly smaller than the spin amplitude ($\delta \ll x_m$) \cite{BOE21}, we take the noise strength to be of the same magnitude as the analog spin amplitude ($\delta \cong x_m$). Similar to the thermodynamic temperature, the noise then acts as a randomizing element for the spins and prevents convergence to a stable state. We conjecture that this can be used to drive analog Ising machines into an equilibrium state at a temperature determined by the noise strength. While noise has been considered in analog Ising machines and digital-analog computing concepts as a way of improving success rates and exploring low energy states \cite{NG22,KAK20,YAM17a,CAI20,ROQ20,PIE20,PRA20}, here we show that noise allows to continuously generate statistically independent samples of the Boltzmann distribution at any arbitrary temperature.

We experimentally demonstrate this noise-induced continuous Boltzmann sampling with a time-multiplexed opto-electronic analog Ising machine. Moreover, in the methods section, we show that noise-induced sampling can also be achieved with other types of analog Ising machines. Our time-multiplexed opto-electronic Ising machine is a hybrid system, which consists of an opto-electronic analog nonlinear system to generate the spin states and a field programmable gated array (FPGA) to perform the spin coupling \cite{BOE19}. Such opto-electronic Ising machines can also be built inexpensively with off-the-shelf components within a compact footprint and are resilient to perturbations. Contrary to spatially-multiplexed Ising machines, the system uses a time-multiplexing scheme with time-discrete feedback, where spins sequentially propagate through the feedback loop in an iterative scheme. This allows to implement hundreds of thousands of spins with a single nonlinear system \cite{HON21}. Fig.\ref{fig1}c shows a schematic of our system. The nonlinear analog system consists of a laser, whose light passes through a Mach-Zehnder modulator (MZM) and is detected by a photodiode. During each iteration $k$, the time-multiplexed sequence of spins is used to modulate the laser intensity and the resulting photovoltage coming from the modulator is digitized by an analog-digital converter (ADC1). The FPGA demultiplexes the signal to obtain the spin amplitudes $x_m$, performs a matrix-vector multiplication and adds biases $b_m$ to generate the feedback signal

\begin{equation}
	f_m[k]=\alpha x_m[k]+\beta \left(\sum_{n}J_{mn}x_n[k]+x_{\mathrm{sat}} b_m\right) \ \ . \label{eq3}
\end{equation}

Here, $\alpha$ is the feedback gain and $\beta$ is the coupling strength. The saturation amplitude $x_{\mathrm{sat}}=\mathrm{max}(\left|x_m\right|)$ is responsible for rescaling the biases in relation to the spin coupling and is determined by the maximal possible spin amplitude, which is internally limited in our experimental system by electrical components to $x_{\mathrm{sat}}=0.7V$. The feedback signal is multiplexed and injected into the input port of the MZM through a digital-analog-converter (DAC) for the next iteration $k+1$ to close the feedback loop. We extend this scheme by adding an additional Gaussian white noise signal with standard deviation $\delta$. A broadband noise signal is generated by an electrical analog noise source that is amplified by an electronic amplifier (EA) with a bandwidth of 300MHz. The noise signal is then digitized by an ADC (ADC2) and added to the feedback signal in equation \eqref{eq3}. We have also compared this analog noise signal against an FPGA-internally generated signal, using a pseudo random number generator, and have verified that both methods result in a comparable evolution of the spin amplitude.

\begin{figure}[htbp]
	\centering
		\includegraphics[width=.99\textwidth]{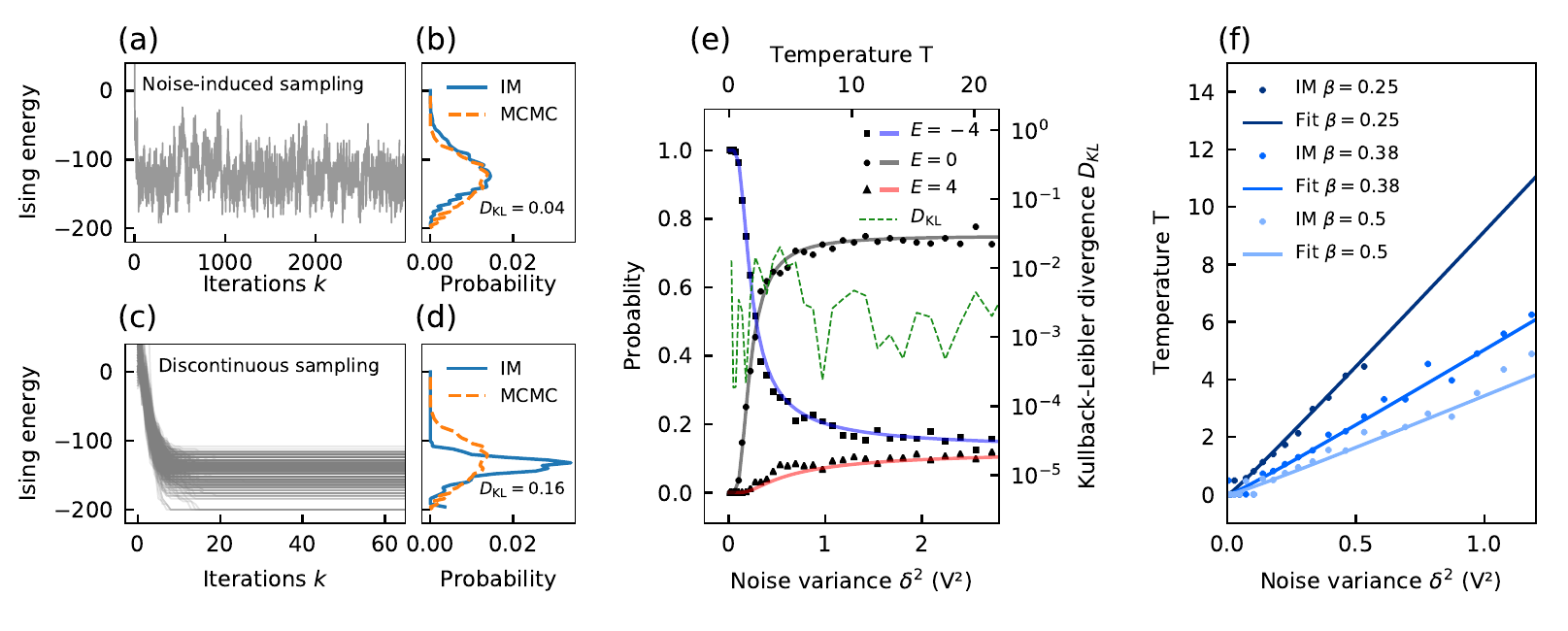}
	\caption{\textbf{Experimental demonstration of noise-induced Boltzmann sampling with a time-multiplexed opto-electronic Ising machine} Time evolution (a,c) and sampled distribution function (b,d) of the Ising energy for noise-induced sampling (a,b) and discontinuous sampling (c,d). In (b,d), the energy distributions obtained with the Ising machine (IM) are compared to those obtained with the Metropolis-Hastings algorithm (MCMC).  (e) Boltzmann distribution obtained from noise-induced sampling as a function of the noise variance $\delta^2$ for the three degenerate energy levels of a 4 spin ring network (dots, squares and triangles) at $\alpha=1.2$ and $\beta=0.5$. The probabilities are compared to the analytical solutions (solid lines) obtained from equation \eqref{eq1} at different temperatures $T$. The sampling overlap of the distribution for the Ising machine is quantified by the Kullback-Leibler divergence $D_\mathrm{KL}$ (dashed line). (f) Relation between temperature and the noise variance for the problem in (e) for different coupling strengths $\beta$.}
	\label{fig2}
\end{figure}

Based on this setup, we experimentally demonstrate that noise-induced sampling is approximating the Boltzmann distribution at a temperature of $T=2$ for an anti-ferromagnetic ($J_{mn}=-1$) square lattice with nearest neighbor coupling consisting of $N=100$ spins. In Fig.\ref{fig2}a, we show the time evolution of the Ising energy for a noise strength of $\delta=1V$ with $\alpha=0.7$ and $\beta=0.5$. Contrary to regular noise-conditions ($\delta \ll x_\mathrm{sat}$), the spin state does not converge to a stable configuration but rather evolves continuously due to the injected noise. In Fig.\ref{fig2}b, we show the resulting approximation of the Boltzmann energy distribution, which is obtained by taking samples after every iteration. We compare a distribution made from 5000 consecutive samples against software-based sampling using Markov chain Monte-Carlo (MCMC) sampling and find that the distribution obtained with the Ising machine agrees very well with software-based sampling. To quantify the sampling accuracy, we employ the Kullback-Leiber divergence $D_{\mathrm{KL}}$, which measures the overlap between two distributions $P_a$ and $P_b$. For discrete probability distributions, it is defined as
\begin{equation}
	D_{\mathrm{KL}}(P_a(x),P_b(x))=\sum_{x \in X} P_a(x)\log\left( \frac{P_a(x)}{P_b(x)} \right) \ \ . \label{eq4_1}
\end{equation}
Here, $x$ are the possible outcomes in the probability space $X$. The Kullback-Leiber divergence is always positive and becomes zero for two perfectly matched distributions, while it diverges for two completely unmatched distributions. Although the Kulback-Leibler divergence is no absolute measure of sampling accuracy, it allows to compare the accuracy between different sampling methods. 

We utilize this to quantify the sampling accuracy compared to existing Ising machine-based sampling methods based on trapping in local energy minima. In Fig.\ref{fig2}c, we show the time evolution for such discontinuous sampling when approximating the same distribution as in Fig.\ref{fig2}b. For this, the Ising machine is initialized 500 times and left to run for 100 iterations until it converges to a stable configuration. The system is operated at a high gain with small noise ($\alpha=1.5$, $\beta=0.5$, $\delta=0.005$), which forces convergence to different high energy states \cite{LEL17,BOE19}. After each run, the last step is taken as a statistical sample and used to estimate the energy distribution. In Fig.\ref{fig2}d, we show the resulting approximated energy distribution and observe a bad overlap with the distribution obtained by MCMC sampling. To compare the sampling accuracy to the noise-induced sampling in Fig.\ref{fig2}b, we calculate the Kullback-Leibler divergence for both distributions in relation to the distribution obtained by MCMC sampling. For the distribution produced with noise-induced sampling in Fig.\ref{fig2}b, we obtain a value of $D_{\mathrm{KL}}\approx0.04$, which is close to a perfect overlap. As we will show in the next section, such accuracy is also comparable to that of MCMC-based sampling for complex large-scale problems. For discontinuous sampling on the other hand, we obtain $D_{\mathrm{KL}}\approx0.16$. Compared to noise-induced sampling, the sampling is therefore significantly less accurate, which makes discontinuous sampling unsuitable for applications where high accuracy is required. An additional advantage over discontinuous sampling is the significantly faster sampling. Discontinuous sampling requires the Ising machine to first reach a stable configuration before a sample can be taken, which creates a significant overhead and drastically reduces the sampling speed. The improved accuracy and the ability to continuously draw samples therefore presents a major advantage of noise-induced sampling.

An additional drawback of discontinuous sampling stems from the difficulty of controlling the temperature $T$. For discontinuous sampling, the temperature is nontrivially linked to the system parameters and complex temperature estimation schemes have to be applied for each specific problem and set of parameters, hence making it difficult to set a temperature a-priori \cite{BEN16}. To derive a general parameter dependence of the temperature for noise-induced sampling, we analyze a simple Ising model for which the Boltzmann distribution in eq.\eqref{eq1} can be calculated analytically. For this, we consider a 4-spin-system, where the spins are ordered in a ring structure and coupled anti-ferromagnetically ($J_{ij}=-1$) to their two nearest neighbors. This system possesses three degenerate energy levels, with the ground state (GS) at ${E_\mathrm{GS}}=-4$ and two excited states (ES) at ${E_\mathrm{ES}}=0$ and ${E_\mathrm{ES}}=4$ (exemplary spin arrangements for each degenerate energy are shown as insets in Fig.\ref{fig2}e). To perform continuous Boltzmann sampling, we let the system run for 1000 iterations at each noise level $\delta$ and take a sample after every iteration to measure the probability of reaching the different energy levels. In Fig.\ref{fig2}e, we show the measured probabilities for the three energy levels as a function of the noise variance $\delta^2$ for $\alpha=1.2$ and $\beta=0.5$. To relate the noise variance to the system's temperature, we analytically calculate the occupation probabilities for the three energy levels as a function of the temperature (solid lines). We observe that, by assuming a linear relationship $T\propto \delta^2$ between temperature and noise variance, a good agreement between the sampled and the calculated probabilities is achieved for the entire temperature range. We also calculate the Kullback-Leibler divergence for the entire temperature range (gray dashed line) and find that it is below $D_{\mathrm{KL}}<0.01$, hence indicating a good overlap to the exact distribution function for each temperature. In Fig.\ref{fig2}f, we show the linear relationship between $T$ and $\delta^2$ for different values for the coupling strength $\beta$. The temperature is extracted from the noise-induced sampling by comparing the sampled and the exact probabilities in Fig.\ref{fig2}e. The relationship between $\delta^2$ and $T$ is then obtained by a linear interpolation. We find that the linear relationship between noise and temperature holds true even as we change the parameters in the system. The temperature follows a linear relationship, where the slope is determined by the coupling strength $\beta$. Contrary to discontinuous sampling, this omits the need to extract nonlinear parameter relationships. Once the slope of the linear relationship has been obtained for a specific problem, the temperature can be controlled a-priori by adjusting the noise power of the injected noise signal. As we show in the following sections, this simple linear relationship also holds true for more general problems as well as for large numbers of spins.

\subsection*{Unsupervised neural network training with analog Ising machines}

\begin{figure}[htbp]
	\centering
		\includegraphics[width=.99\textwidth]{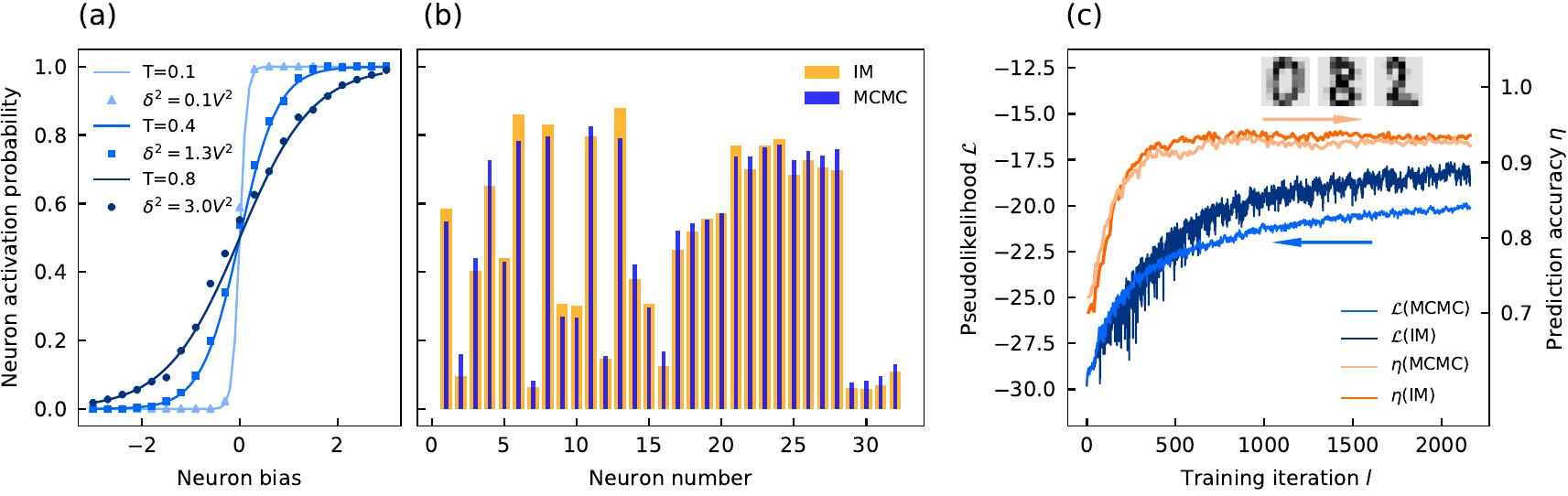}
	\caption{\textbf{Experimental demonstration of unsupervised training of RBMs with a time-multiplexed opto-electronic Ising machine} (a) Activation probability of single neurons as a function of the neuron bias for different temperatures. The probabilities have been obtained from continuous sampling of 100 independent Ising spins at different noise levels (squares, dots and triangles) and are compared to the analytical solution at different temperatures (solid lines). (b) Activation probabilities for an RBM with 16 hidden and 16 visible neurons with random weights and biases. Probabilities for the analog Ising machine (IM, orange bars) have been obtained by continuous sampling at a fixed noise strength and are compared against probabilities obtained with the Metropolis-Hastings algorithm (MCMC, blue bars). (c) Comparison of the pseudolikelihood $\mathcal{L}$ and the prediction accuracy $\eta$ for Ising machine- and MCMC-based sampling during unsupervised training of handwritten digit recognition.}
	\label{fig3}
\end{figure}

With its ability to sample Boltzmann distributions, we want to experimentally investigate the application of noise-induced sampling with our time-multiplexed opto-electronic Ising machine to the training of neural networks. In the methods section, we also demonstrate how this can be extended to other kinds of analog Ising machines. Boltzmann sampling is required in the unsupervised training of a variety of stochastic generative neural network architectures, s.c. Boltzmann machines, whose energy function can be directly mapped to the Ising Hamiltonian \cite{ACK85,MUN11}. During training, the weights and biases of the neurons are optimized in a gradient descent with the learning rate $\epsilon$ to maximize the logarithmic likelihood between the neural network's output and the training data (details about the mapping and the learning algorithm are given in the methods section) \cite{ACK85}. For each update, Boltzmann sampling of the neuron activation probabilities has to be performed using MCMC-based sampling, which is known to be computationally difficult \cite{LON10}. Here, we experimentally demonstrate noise-induced sampling running on the opto-electronic Ising machine in Fig.\ref{fig1}c to replace software-based sampling in the training of Boltzmann machines. We first consider emulating the behavior of single neurons of a Boltzmann machine. For Boltzmann machines, the activation probability of a neuron as a function of its input is given by a logistic function \cite{MUN11}. In Fig.\ref{fig3}a, we emulate single neurons of a Boltzmann machine on the opto-electronic Ising machine, by taking individual single Ising spins and sampling their activation probability through continuous sampling for 1000 iterations at $\alpha=1.2$. Samples are obtained after every iteration $k$. To obtain the activation probability function, we average the spin state over all iterations as a function of the neuron bias for different noise levels averaged over 100 spins. When fitting the activation probability with logistic functions at different temperatures, we find a good agreement between the analog Ising machine and the analytical solution. As before, the temperature can be directly adjusted through the noise strength injected into the analog Ising machine. We also consider sampling of the activation probabilities within arbitrary neural networks. In Fig.\ref{fig3}b, we measure the activation probabilities for a neural network consisting of 32 neurons with random weights and spin biases at a noise level corresponding to a temperature of ${T}=1$. The activation probabilities of the individual neurons are obtained by performing continuous sampling for 20000 iterations at $\alpha=0.75$, $\beta=0.2$ and $\delta=0.3V$ by taking samples after every iteration $k$. We compare the probabilities to those obtained through software-based sampling using the Metropolis-Hastings algorithm and find a good agreement between the analog Ising machine and the software-based sampling, hence indicating that sampling in Boltzmann machines can be successfully emulated with analog Ising machines. 

Finally, we use our time-multiplexed Ising machine to experimentally demonstrate unsupervised training in an image recognition task. In Fig.\ref{fig3}c, we show the unsupervised training for handwritten digits with a restricted Boltzmann machine (RBM). An RBM is a neural network consisting of a single visible and a single hidden layer. The nodes between the layers are fully connected, while there are no intra-layer connections. For the handwritten recognition task, we consider a single RBM layer consisting of 100 hidden and 64 visible neurons, which is trained with a learning rate of $\epsilon=0.2$. The training dataset contains 7188 8 by 8 greyscale images of single handwritten digits \cite{ALI97} (sample images are shown as insets in Fig.\ref{fig3}c). During each training iteration $l$, a minibatch of 100 training images is used for training and the activation probabilities are approximated with the analog Ising machine from 1000 samples. Samples are obtained after every iteration of the analog Ising machine. The feedback, coupling and noise strength are left constant for each training iteration $l$ at $\alpha=0.5$ and $\beta=0.75$ and $\delta=0.6V$. The constant parameters are an important advantage over discontinuous sampling, where the sampling temperature can shift during the training as neuron weights are updated \cite{BEN16}. As the training of Boltzmann machines is known to be very sensitive to the temperature, discontinuous sampling requires a regular re-estimation of the temperature and frequent adjustments of the parameters to maintain accurate sampling. 

We track the progress of the training as a function of the training iteration $l$ by analyzing the pseudolikelihood $\mathcal{L}$, which is a common measure in machine learning and indicates how closely the RBM is approximating the training data \cite{BES75}. When initializing the RBM with random weights, we observe a quick increase of the pseudolikelihood with the training before it begins to saturate after $l\approx 2000$ training iterations. We compare the Ising machine-based sampling to training using MCMC-based sampling with the iterative Metropolis-Hastings algorithm. During each training step, we run the Metropolis-Hastings algorithm for 20000 iterations and samples are obtained after every iteration. When comparing the two methods, we observe a very similar evolution of $\mathcal{L}$ with an overall higher maximum pseudolikelihood achieved by the analog Ising machine ($\mathcal{L}_\mathrm{IM}=-17.6$, $\mathcal{L}_\mathrm{MCMC}=-19.9$). This indicates that the Ising machine is able to achieve a better approximation of the training data. To analyze the impact of this on the recognition accuracy of the handwritten digits, we combine the RBM with a single logistic regression layer. The regression layer is trained on the activation probabilities of the hidden neurons in a supervised way after each unsupervised training step of the RBM. The prediction accuracy $\eta$ for recognizing individual digits is shown as a function of $l$ in Fig.\ref{fig3}c. After $l\approx300$ training iterations, the prediction accuracy saturates at its maximum value $\eta_{\mathrm{IM}}=0.943$. Compared to just a single prediction layer without an RBM ($\eta=0.77$), the prediction accuracy is significantly improved by using an RBM. The accuracy of Ising machine-based training is comparable to training performed by MCMC-based sampling, with a slightly lower maximal accuracy achieved by the MCMC sampler ($\eta_{\mathrm{MCMC}}$=0.938). The accuracy is also comparable to that of training the RBM with the contrastive divergence algorithm. Contrastive divergence is an approximation of the probability distribution and is commonly used in the training of RBMs \cite{CAR05}. It achieves a maximal prediction accuracy of $\eta_{\mathrm{CD}}=0.948$ and a maximal pseudolikelihood of $\mathcal{L}_\mathrm{CD}=-19.9$. Overall, the handwritten digit recognition task demonstrates that noise-induced sampling with analog Ising machines can be successfully used to train RBMs. Remarkably, our experimental setup achieves a slightly improved performance compared to MCMC-based sampling on digital computers and shows that software-based sampling can be replaced by analog Ising machines. 

\subsection*{Scalability and speed of analog Ising machine-based Boltzmann sampling}

\begin{figure}[htbp]
	\centering
		\includegraphics[width=.99\textwidth]{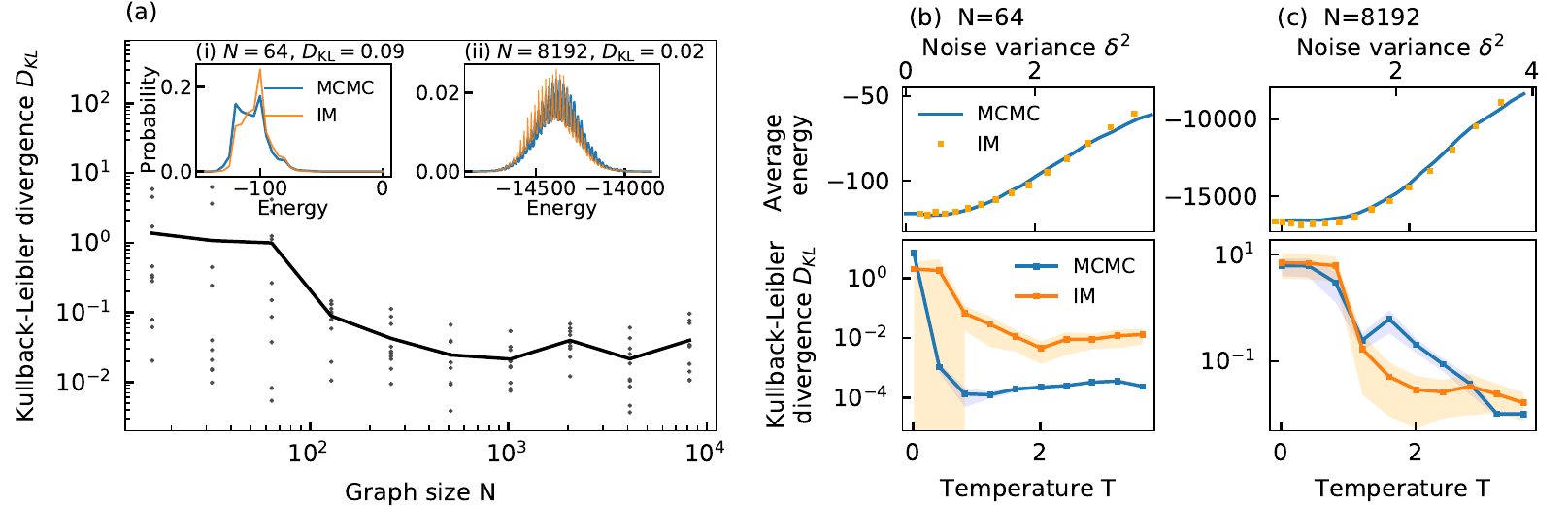}
	\caption{\textbf{Simulation-based investigation of the scalability of noise-induced Boltzmann sampling} (a) Kullback-Leibler divergence as a function of the problem size for sampling the energy distributions for different random sparse graphs (dots). The solid line shows the average. The insets (i) and (ii) show exemplary sampled distributions in comparison to the Metropolis-Hastings algorithm (MCMC) for $N=64$ (i) and $N=8192$ (ii) spins. (b) and (c) Average energy (top panel) and Kullback Leiber divergence (lower panel) as a function of the temperature and the noise variance $\delta^2$ for a sparse random graph with $N=64$ (b) and $N=8192$ (c). The average energy and Kullback-Leiber divergence are compared against MCMC-based sampling (blue lines). For $D_\mathrm{KL}$ for the MCMC-based sampling, repeated sampling runs are compared against the reference distribution. The shaded regions show the standard deviation.}
	\label{fig4}
\end{figure}

Beyond the sampling problems in Fig.\ref{fig2} and Fig.\ref{fig3} with $N\leq 164$, we also demonstrate the scalability of Ising machine-based sampling to complex large-scale problems with $N\leq8192$. Of particular interest is the scaling of the sampling accuracy and the sampling speed for large spin numbers and varying temperatures. Both are known to increase the complexity of the sampling for MCMC-based methods, i.e. due to a large probability space and effects such as critical slowing down \cite{ARO18}. To probe the scalability of Ising machine-based sampling, we perform sampling on a set of randomly generated graphs with antiferromagnetic coupling ($J_{mn}=-1$) for spin numbers from $N=16$ to $N=8192$. Such random graphs are of particular interest, as they are known to present difficult benchmark instances for combinatorial optimization problems \cite{KAL20}. For each problem size $N$, we create 10 different graphs, where each node is randomly coupled to 8 other nodes in the graph. To study these graphs, we perform numerical simulations of a spatially-multiplexed analog Ising machine, as shown in Fig.\ref{fig1}a. We chose this simulation-based approach, as it enables us to study problem sizes beyond the current capabilities of our experimental system. Moreover, it allows us to generalize the performance of noise-induced sampling to different realizations of analog Ising machines, so that we can demonstrate the universality of our noise-induced sampling approach. The numerical model we employ is a more general time-continuous description of the spatially-multiplexed analog Ising machine in Fig.\ref{fig1}a \cite{BOE21}. Compared to our time-multiplexed experimental system in Fig.\ref{fig1}, time-continuous analog Ising machines implement the Ising model in a fully analog way \cite{TEZ20,CHO19,HON21,MOY22}. Here, spin state generation and spin coupling are performed in parallel, e.g. using MZM networks or resistor-crossbar arrays. This presents a highly efficient implementation of Ising machines, as such a system does not suffer from slowdown due to temporal multiplexing, analog-digital conversion and the processing speed of an FPGA. In principal, the speed of such a fully analog system is primarily limited by the analog bandwidth of the components, which can be in the order of tens of gigahertz for opto-electronic and all-optical systems \cite{TIA18,CHO19,TEZ20,MOY22}. This makes analog Ising machines quite different from analog-digital computing approaches \cite{CAI20,PIE20,PIE19,PRA20,ROQ20}. Such systems are software-based implementations of the Ising model, similar to simulated annealing and Hopfield networks \cite{HOP85,KIR83}, that utilize large-scale analog vector-matrix multipliers to accelerate computationally expensive calculations. Contrary to a fully analog computing system, they require digital computers to store and process the spin states and are therefore bound by the clock frequencies of their digital processing units. It is important to note that our experimental time-multiplexed system in eq.\eqref{eq3} is equivalent to a numerical integration of the equations of motion of a spatially-multiplexed analog Ising machine \cite{BOE20}. As we show in the methods sections, a time-multiplexed system thus presents a good approximation of a spatially-multiplexed analog Ising machine, so that findings from the numerical simulations can also be generalized to our experimental system.

To quantify the sampling accuracy, we compare the distributions obtained by noise-induced sampling with an analog Ising machine against a reference distribution obtained by MCMC-based sampling at a fixed temperature of $T=2$. For the reference distribution, $P_a$ in eq.\eqref{eq4_1} is approximated with the iterative Metropolis Hastings algorithm. For each graph, the distribution is approximated from 20 million sampling steps to ensure an accurate reference distribution, as the Metropolis-Hastings algorithm is known to converge to the true distribution in Eq.\eqref{eq1} for a large number of samples \cite{HOL98}. Here, samples are obtained after every 100th sampling step. For the sampling with the analog Ising machine, $P_b$ in eq.\eqref{eq4_1} is approximated from 100,000 samples, where a sample is taken after each single simulation iteration. For each random graph, the noise level $\delta$ is adjusted accordingly to achieve a good overlap with the reference distribution. The feedback strength and the coupling strength are fixed at $\alpha=0.3$ and $\beta=0.2$, although sampling is also possible with other parameter combinations by adjusting the noise level (see Fig.\ref{fig2}f). In Fig.\ref{fig4}a, we show the resulting Kullback-Leibler divergence when comparing Ising machine-based sampling against the MCMC-based reference distribution. On average, we find that Ising machine-based sampling is able to achieve a Kullback-Leibler divergence well below $D_{\mathrm{KL}}<0.1$ for problem sizes beyond $N\approx 128$ and therefore close to a perfect overlap. As a reference, the insets (i) and (ii) in Fig.\ref{fig4}a show sampled distributions in comparison to MCMC-based sampling for $N=64$ (i) and $N=8192$ (ii). For the sampled distributions, the Kullback-Leibler divergence is around $D_{\mathrm{KL,64}} \approx 0.09$ and $D_{\mathrm{KL,8192}} \approx 0.02$ and we observe a good match to the MCMC-sampled reference distributions. For smaller problem sizes, we observe a rise in the average Kullback-Leibler divergence, as some individual problems become harder to sample accurately with the analog Ising machine. For these smaller problems, we note that there is a high probability for the system to be in the ground state. This can become problematic for analog Ising machines, as the inhomogenous analog spin amplitude can induce mapping errors to the Ising Hamiltonian. In these cases, the analog Ising machine has a lower probability than expected for reaching the ground state, so that accurate sampling of them becomes more difficult \cite{LEL17,BOE21}. Interestingly, we find that this issue does not exist for larger spin numbers, so that accurate sampling with analog Ising machines is scalable to large-scale complex problems.

We also study the temperature $T$ in Fig.\ref{fig4}b and c to establish its influence on the sampling accuracy and to demonstrate that the linear noise-temperature relationship in Fig.\ref{fig2}e holds true for more general problems. In Fig.\ref{fig4}b, we study an exemplary graph from Fig.\ref{fig4}a for $N=64$. In the top panel, we show the average energy as a function of the temperature as obtained from MCMC-based sampling and compare it to noise induced sampling with the analog Ising machine. For the analog Ising machine, the noise variance $\delta^2$ is swept in a similar fashion to Fig.\ref{fig2}e for $\alpha=0.9$ and $\beta=0.1$. As in Fig.\ref{fig2}e, a linear relation between $\delta^2$ and $T$ is shown by sweeping $\delta^2$ and comparing the average energy of the sampled distribution to that of MCMC-based sampling at different temperatures. As with the 4-spin problem, we find that a linear sweep of $\delta^2$ can very well reproduce the energy-temperature relation of MCMC-based sampling for a large temperature range. This indicates that the a-priori control of the temperature is also possible for more general problems. In the lower panel, we show the Kullback-Leibler divergence for the different temperatures when comparing the distribution obtained with noise-induced sampling to an MCMC-sampled reference distribution. Averaged over 10 independently sampled distributions, we observe that, while $D_{\mathrm{KL}} < 0.1$ for temperatures above $T\gtrsim 1$, the sampling accuracy of the analog Ising machine significantly deteriorates for lower temperatures. This is expected, as trapping in local energy minima becomes more prevalent at low temperatures. Here, it becomes easy to end up in different energy minima for different initial conditions, so that sampled distributions can differ significantly between repeated runs. It is important to note that such temperature-dependent trapping in local energy minima is a common phenomenon in MCMC methods. For higher temperatures, $D_{\mathrm{KL}}$ reaches a minimum of $D_{\mathrm{KL}} \approx 0.01$ and shows that accurate sampling with the Ising machine is possible over a large temperature range.

Finally, we provide a more direct comparison of the sampling accuracy between noise-induce sampling and MCMC-based sampling with the Metropolis-Hastings algorithm. For this, MCMC-sampling is compared against its own reference distribution by performing repeated sampling of the same graph. Here, the same number of iterations is evaluated as for the reference distribution. Due to the non-deterministic nature of MCMC-based sampling, such repeated runs will result in small deviations between the sampled distributions, which we quantify with the Kullback-Leibler divergence. In the lower panel of Fig.\ref{fig4}b, we show the Kullback-Leibler divergence for 10 repeated sampling runs at different temperatures and observe that, between the different runs, there can be significant differences in the sampled distribution. For low temperatures in particular, this results in a significant increase in the Kullback-Leibler divergence. Here, the sampling accuracy is on a very similar level to Ising machine-based sampling. For increasing temperatures, the overall accuracy of the MCMC-based sampling increases and achieves a Kullback-Leiber divergence of $D_{\mathrm{KL}}\approx 0.0001$. For the small-scale graph considered here, while the accuracy of Ising machine-based sampling can already be quite high, the Metropolis-Hastings algorithm can achieve a further improvement in accuracy. We perform the same analysis as in Fig.\ref{fig4}b for a large-scale graph. In Fig.\ref{fig4}c, we show an exemplary graph from Fig.\ref{fig4}a with $N=8192$ spins. When comparing the average energy between MCMC-based and Ising-machine based sampling in the upper panel, we again observe a good agreement in the energy-temperature relationship. As for the small-scale graph in Fig.\ref{fig4}b, this indicates that the linear relationship between temperature and noise variance also holds true for large spin numbers. For the Kullback-Leibler divergence in the lower panel, we find that the sampling accuracy is quite comparable between Ising machine-based and MCMC-based sampling. Remarkably, we find that $D_{\mathrm{KL}}$ for the Ising machine-based sampling is lower compared to MCMC-based sampling for temperatures between $T\gtrsim1$ and $T\lesssim3$, thus showing that the sampling accuracy is on-par or even improved compared to the Metropolis-Hastings algorithm for this large-scale graph. 

\begin{figure}[htbp]
	\centering
		\includegraphics[width=.49\textwidth]{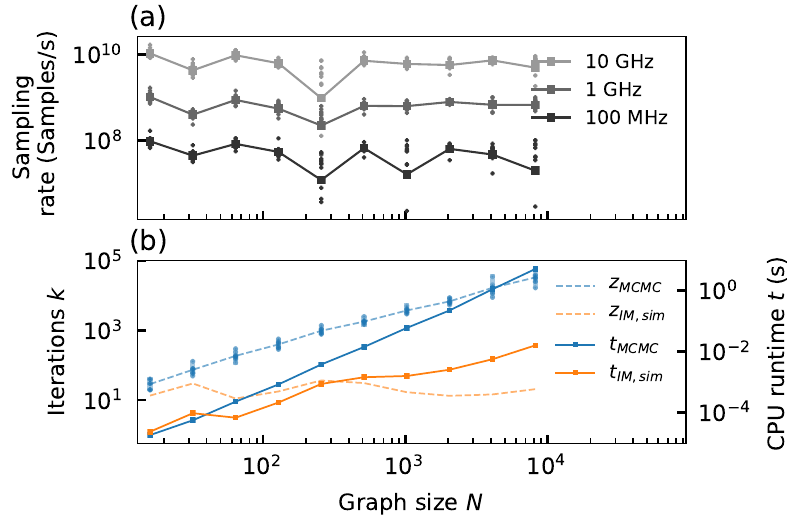}
	\caption{\textbf{Simulated sampling rate of spatially-multiplexed analog Ising machines} (a) Estimation of the sampling rate of a spatially-multiplexed analog Ising machine at different analog bandwidths for randomly generated sparse Maxcut problems at a temperature of $T=2$ (dots). The lines indicate the average for the different graphs. The sampling rate is estimated from the autocorrelation function of the Ising energy as the point when samples become statistically independent. (b) Number of iterations required $z$ to create statistically independent samples with the Metropolis-Hastings algorithm (MCMC) and with simulations of analog Ising machines (IM,sim) using the forward Euler method. Also shown is the average runtime $t$ to obtain independent samples when executing the Metropolis-Hastings algorithm and the forward Euler integration on the same CPU.}
	\label{fig5}
\end{figure}

We also analyze the scalability of the sampling time for Ising machine-based Boltzmann sampling. In Fig.\ref{fig5}a, we estimate the potential sampling rate of a spatially-multiplexed analog Ising machine in sampling the energy distributions in Fig.\ref{fig4}a at a temperature of $T=2$ for different bandwidths $B=1/(2\pi\tau_l)$. To determine the sampling rate $1/\tau_{\mathrm{cor},B}$ at bandwidth $B$, we calculate the correlation time $\tau_{\mathrm{cor}}$ from the autocorrelaton function $C_{\mathrm{auto}}(\tau)=\frac{\left\langle(E(t)-\bar{E})(E(t-\tau)-\bar{E})\right\rangle}{\langle(E(t)-\bar{E})(E(t)-\bar{E})\rangle}$, where $\left\langle ...\right\rangle$ denotes a time average and $\bar{E}$ is the average energy. $\tau_{\mathrm{cor},B}$ corresponds to the value where $C(\tau)$ has decayed by $1/e$ and provides an estimate for the minimum time period between statistically independent samples. We can therefore use the autocorrelation time as an estimate for the maximally possible sampling rate. Based on this, in Fig.\ref{fig5}a, we analyze the sampling rate for Ising machines with analog bandwidths ranging from 10 GHz to 100 MHz and observe a linear scaling of the sampling rate with the bandwidth. Importantly, we note that the average sampling rate is not affected by the problem size. This is due to the simultaneous injection of noise into all spin systems. Compared to the iterative Metropolis-Hastings algorithm, this enables very efficient mixing so that fast exploration of different energy states is possible. For the largest problem size ($N=8192$), the average simulated sampling rate is then at $1/\tau_{\mathrm{cor,10GHz}}=4.9\mathrm{GS/s}$, $1/\tau_{\mathrm{cor,1GHz}}=667\mathrm{MS/s}$ and $1/\tau_{\mathrm{cor,100MHz}}=19\mathrm{MS/s}$. 

To compare the speed of the analog system to software-based sampling, we consider the sampling rate of MCMC-based sampling using the iterative Metropolis-Hastings algorithm. Similar to the sampling rate for the analog Ising machine, in Fig.\ref{fig5}b, we calculate the number of sampling steps $z_\mathrm{MCMC}$ required to generate statistically independent samples of the energy distribution. We observe an exponential scaling of the number of iterations with the problem size $N$, which is a common feature and a major contributor to the computational complexity of MCMC-based sampling \cite{ARO18}. For the largest problem size in Fig.\ref{fig5}b ($N=8192$), $z_{\mathrm{MCMC}}\approx 30,000$ sampling steps with the Metropolis-Hastings algorithm are required to obtain statistically independent samples. For highly efficient digital implementations of the Metropolis Hastings algorithm running on parallel FPGAs, where iterations can be performed within approximately 10 picoseconds \cite{BEL07}, this would translate to an average sampling rate that is six times slower than an analog Ising machine with a bandwidth of 100MHz. With a 10 GHz analog Ising machine, sampling rates can then already be improved by a factor of 1000 over software-based sampling. With the ability to scale analog Ising machines to much higher bandwidths, implementing noise-induced sampling on analog Ising machines therefore offers a potential orders-of-magnitudes speedup in the sampling speed, while the sampling quality is comparable to MCMC-based sampling. 

Interestingly, we find that even software simulations of noise-induced sampling can be more efficient than the Metropolis-Hastings algorithm. In Fig.\ref{fig5}b, we show the number of iterations $z_{\mathrm{IM,sim}}$ that are required during our simulations for achieving independent samples with analog Ising machines. Compared to the Metropolis-Hastings algorithm, the number of iterations does not increase exponentially with the problem size. Especially for large problems, we find that the number of iterations can be more than 100 times smaller compared to the MCMC-based sampling. Based on the number of iterations, we determine the average runtime of the Ising machine simulations $t_\mathrm{IM,sim}$ and of the Metropolis-Hastings algorithm $t_\mathrm{MCMC}$ for obtaining statistically independent samples. For this, both algorithms are executed on the same CPU. Overall, we find that due to the lower number of iterations, the overall runtime is significantly faster for the Ising machine simulations. For the largest problem sizes, we measure up to 300 times shorter computation times for the Ising machine simulation. While not as fast as a fully analog realization of an Ising machine, this also makes software-based implementations of analog Ising machines a promising alternative to MCMC-based sampling.


\subsection*{Discussion}

We find that noise-induced sampling presents a significant improvement in speed and ease-of-use for Boltzmann sampling with analog Ising machines. While current discontinuous sampling requires the analog Ising machine to perform a full initialization and equilibration cycle to obtain just a single sample, noise-induced sampling omits these bottlenecks and allows to generate samples at rates close to the bandwidth of the analog system. Compared to discontinuous sampling running on quantum annealers, where samples can be generated at rates of around 7 kS/s \cite{LIU18}, order-of-magnitudes improvements in the sampling speed are possible with an analog Ising machine. The concept of injecting noise from an analog source can also be adapted to other optical or electronic systems (see methods section) and makes noise-induced sampling a universal concept for Ising machine-based sampling. Our method can easily be applied to existing large-scale analog Ising machines, e.g. based on optical parametric or electronic oscillators \cite{HON21,MOY22}, to extend their range of applications. By leveraging the inherently high analog bandwidth of such systems, this can lead to a generation of highly efficient statistical samplers that operate orders-of-magnitude faster compared to digital computers. While not as fast as a fully analog realization, noise-induced sampling can also be adapted into Ising-machine inspired sampling algorithms. Similar to Ising machine-inspired optimization algorithms \cite{GOT21}, the reduced number of computations compared to the Metropolis-Hastings algorithm provides a promising route to increase the efficiency of Boltzmann sampling, similarly to other metaheuristic approaches such as cluster algorithms \cite{SWE87,WOL89}. Furthermore, noise-induced sampling could also serve as a way for augmenting other metaheuristic approaches, .e.g., as fast samplers for parallel tempering \cite{SWE86}. 

The improvements in sampling efficiency provided by noise-induced sampling can yield significant advantages for a number of applications. As Boltzmann sampling presents the most computationally expensive task in training Boltzmann machines, we expect the improvements in the sampling speed of analog Ising machines to translate into significantly reduced training times, thereby bridging the existing efficiency gap in applying analog Ising machines to machine learning tasks. Compared to approximate software-based training methods, such as contrastive divergence, full Boltzmann sampling is guaranteed to converge to the correct distribution \cite{CAR05,HOL98}. For more complex tasks and ambiguous input data, as well as for more general types of Boltzmann machines, a higher performance can therefore be expected, while contrastive divergence can introduce unwanted biases to the training \cite{CAR05,FIS11}. Beyond restricted Boltzmann machines, noise-induced sampling can also enable effective training of more general neural network structures, which are not accessible with approximate training methods. As Boltzmann sampling is ubiquitous in various other applications, we also see a large potential for using analog Ising machines in other fields, such as finance or drug research. Analog Ising machines using discontinuous sampling have already demonstrated their use in structure based-screening for drug development \cite{SAK16}. With the improved accuracy and higher speed of noise-induced sampling, analog Ising machines can provide acceleration for drug design. Noise-induced sampling therefore extends the usability of analog Ising machines as accelerators for machine learning and opens up applications beyond combinatorial optimization. 

\section*{Methods}

\subsection*{Time-multiplexed opto-electronic Ising machine}

The time-multiplexed opto-electronic Ising machine uses a nonlinear optical system, consisting of a laser, a Mach-Zehnder intensity modulator and a photodiode. The laser is a single-mode wavelength-stabilized DFB laser diode at a wavelength of $\lambda=1.55 \mu m$. The laser is operated at approximately two times its threshold current at an optical power of around 0.3 mW. The optical modulator is a lithium niobate MZM with an analog bandwidth of 13GHz. A constant bias of $V_{bias}=3.5V$ is applied to the modulator, which corresponds to half of its $V_\pi$ voltage. In the experiments, the sign of $\alpha$ and $\beta$ are inverted, since the transfer function of the MZM has a negative slope at this point. To relate the values of $\alpha$ and $\beta$ in the experiments to the simulations, both $\alpha$ and $\beta$ are rescaled to feedback strength level, where the bifurcation point of an uncoupled system occurs. For both the numerical model and the experimental system, the bifurcation is then at $\alpha=1$. The signal coming from the modulator is detected with a 150 MHz bandwidth GaAs photodiode. For the data acquisition and processing of the spin states, we employ an FPGA system. During each iteration, a network of $N$ spins is generated by time-multiplexing, where the acquisition time is divided into N intervals. The feedback signal $f_n [k]$ is encoded in a piecewise constant function, where the amplitude of each interval is equivalent to the spin amplitude $x_n$. Each interval is 7 $\mu s$ long, which results in an effective sampling rate of 139,000 spins per second. The signal is generated by the FPGA and converted into an analog waveform by a 14-bit DAC before being injected into the input port of the MZM. Due to the voltage limitations of the ADC, the signal is limited to $\pm 0.7V$. The signal coming from the photodiode is simultaneously digitized by a 14-bit ADC and the spin states are extracted by subtracting the DC bias of the photodiode and sampling the amplitude of the signal. The FPGA then performs the matrix-vector multiplication in equation \eqref{eq3} to generate the signal for the next iteration. To inject the noise signal, we generate a Gaussian white noise signal using the built-in noise source of a Tektronix AWG520 arbitrary waveform generator. The signal is amplified by a 300 MHz operational amplifier before being digitized by an ADC. The signal is then added to the feedback signal $f_n [k]$ by the FPGA, with the noise strength being adjusted numerically in the FPGA. We have compared this analog noise source against an FPGA-internally generated signal, using a pseudo random number generator, and have verified that the accuracy of the noise-induced sampling is equal between the two different methods. For a fully analog Ising machine, it is therefore feasible to use analog noise sources to achieve noise-induced sampling.

\subsection*{Numerical model}

Spatially-multiplexed opto-electronic Ising machines are time-continuous feedback systems, whose time evolution is modeled by the following ordinary differential equation \cite{BOE21}:

\begin{equation}
\frac{dx_m}{dt}=\frac{1}{\tau_l}  \left(-x_m (t)+\cos^2\left[(\alpha x_m (t)+\beta \sum_n(J_{mn} x_n (t)+x_\mathrm{sat} b_m )+\delta\zeta(t)-\pi/4\right]-1/2 \right)\ \ .\label{eq_s1}       
\end{equation}

\begin{figure}[htbp]
	\centering
		\includegraphics[width=.99\textwidth]{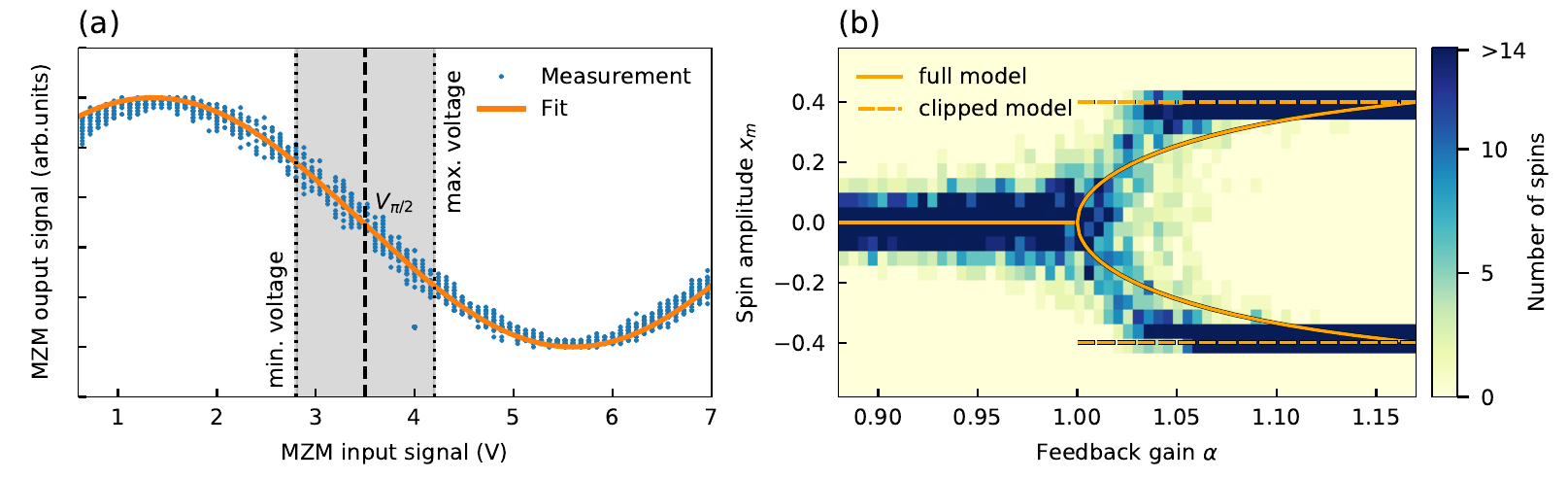}
	\caption{\textbf{Characterization of the time-multiplexed opto-electronic Ising machine} (a) Measurement of the MZM output as a function of the input voltage. The dashed line shows the operating bias voltage of the Ising machine and the grey region denotes the voltage range. (b) Measurement of the spin amplitude distribution as a function of the gain for 100 uncoupled spins ($\beta=0$). The fixed points are compared against the numerical models in equation \eqref{eq_s1} and equation \eqref{eq_s2}.}
	\label{fig6}
\end{figure}

In the following, we refer to eq.\eqref{eq_s1} as the full model. The system is bandwidth limited by a low-pass filter at the frequency $B=1/(2\pi\tau_l)$. Noise is modeled by a Gaussian white noise term $\delta\zeta(t)$ with zero mean and a standard deviation of $\delta$. To drive the system to the correct operating point, a bias of $-\pi/4$ is applied, which corresponds to the point where half of the optical input power passes through the MZM. The transfer function of the MZM is modeled by a $\cos^2$ nonlinearity and a constant of 1/2 is subtracted to remove the DC bias. To better approximate the behavior of the time-multiplexed Ising machine in the experiments, the above model is modified. In the experimental setup (Fig.\ref{fig1}c), a crucial aspect arises due to the internal voltage limitations, which limit the feedback signal to a maximal amplitude of $\pm 0.7V$. Fig.\ref{fig6}a shows the transfer function of the MZM as a function of the input voltage. The system is biased at around 3.5V, which corresponds to the $V_{\pi/2}$ point. Around this bias, the feedback signal is able to modulate the MZM output within a limited voltage range depicted by the grey region. We find that this modulation is significantly smaller than $V_\pi$, so that the transfer function is effectively almost linear within the modulation range. Furthermore, the voltage limitations clip the amplitude to the maximum and minimum voltages for large gain. To account for this behavior, we model the analog Ising machine with the following ordinary differential equation:

\begin{equation}
\frac{dx_m}{dt} = \left\{
\begin{array}{ll}
\frac{1}{\tau_l}\left((\alpha-1)x_m + \beta(\sum_n{J_{mn}x_n}+x_\mathrm{sat}b_m) + \delta\zeta(t)\right) & \left|x_m\right| \leq 0.4 \\
0 & \, \textrm{else} \\ 
\end{array}
\right. \label{eq_s2}
\end{equation}

In the following, we refer to eq.\eqref{eq_s2} as the clipped model. In Fig.\ref{fig6}b, we experimentally measure the spin amplitude distribution of the time-multiplexed Ising machine for 100 uncoupled spins ($\beta=0$) as a function of $\alpha$. We compare this distribution with the fixed points obtained from the full model and the clipped model. While for a gain close to the bifurcation point ($\alpha=1$), the experimental system is close to the full model in equation \eqref{eq_s1}, for large gain, the clipped model better matches the behavior of the experimental system. As we are operating at a gain, that is further away from the bifurcation point, we have therefore selected the clipped model for the simulations of analog opto-electronic Ising machines.

It is important to note that the temporal evolution of our time-multiplexed opto-electronic setup is equivalent to a numerical integration of the time-continuous model in eq.\eqref{eq_s2}. This can be seen by considering a forward Euler integration of eq.\eqref{eq_s2}. The forward Euler integration is a standard method of approximating partial differential equations $\dot{x}(t)=f(x(t))$ and works by discretizing the continuous time $t$ into small time steps. For each time step, the time evolution of $x(t)$ is then approximated from the previous time step in an iterative way:

\begin{equation}
	x[k+1]=x[k]+h\ f(x[k]) \ \ . \label{eq_s22}
\end{equation}

For every iteration $k$, $x(t)$ is then well approximated for sufficiently small step widths $h$. Inserting the equation of motion of the Ising machine in eq.\eqref{eq_s2} into eq.\eqref{eq_s22}, we obtain:
\begin{equation}
x_m[k+1] = \left\{
\begin{array}{ll}
x_m[k]+\frac{h}{\tau_l}\left((\alpha-1)x_m[k] + \beta(\sum_n{J_{mn}x_n[k]}+x_\mathrm{sat}b_m) + \delta\zeta(t)\right) & \left|x_m[k+1]\right| \leq 0.4 \\
x_[k] & \, \textrm{else} \\ 
\end{array}
\right. \label{eq_s23}
\end{equation}

We note that for $h=\tau_l$, the equation simplifies to 

\begin{equation}
x_m[k+1] = \left\{
\begin{array}{ll}
\alpha x_m[k] + \beta(\sum_n{J_{mn}x_n[k]}+x_\mathrm{sat}b_m) + \delta\zeta(t) & \left|x_m[k+1]\right| \leq 0.4 \\
x_m[k] & \, \textrm{else} \\ 
\end{array}
\right. \label{eq_s24}
\end{equation}

\begin{figure}[htbp]
	\centering
		\includegraphics[width=.99\textwidth]{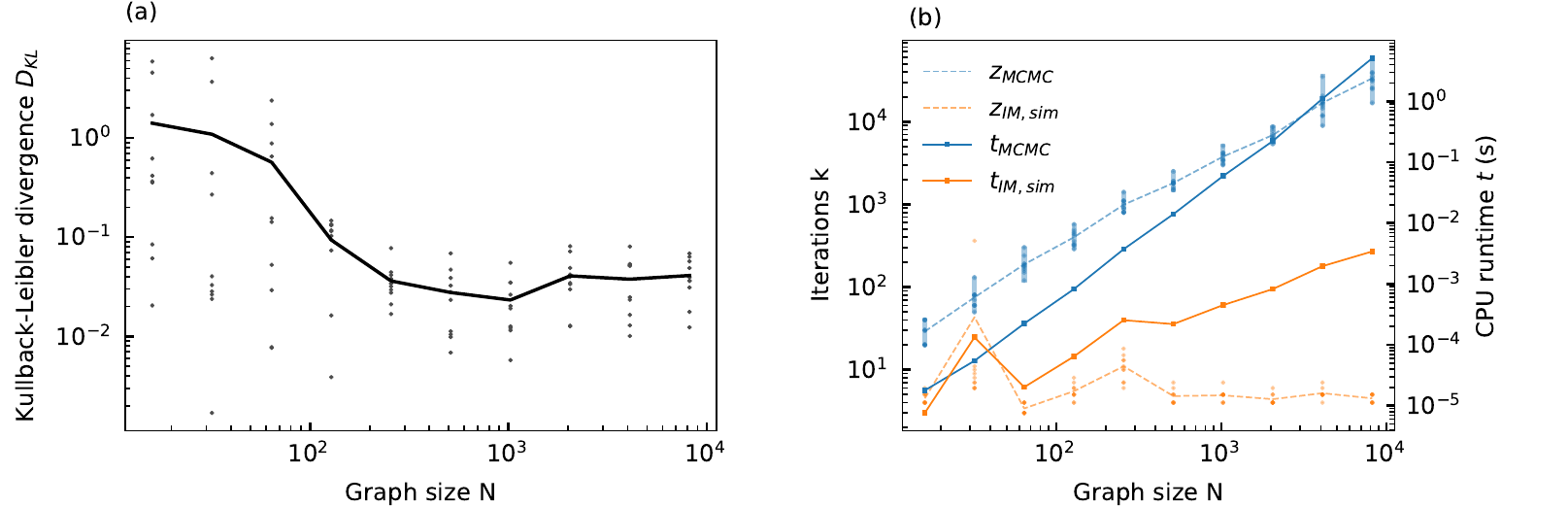}
	\caption{\textbf{Scalability of sampling accuracy and speed for time-multiplexed analog Ising machines} (a) Kullback-Leibler divergence as a function of the problem size for sampling the energy distributions for the different random sparse graphs in Fig.\ref{fig4}a (dots) using simulations of the time discrete model in eq.\eqref{eq_s24}. The solid line shows the average. (b) Number of iterations $z$ required to create statistically independent samples for the graphs in (a) with the Metropolis-Hastings algorithm (MCMC) and with simulations of analog Ising machines (IM,sim) using the time-discrete model. Also shown is the average runtime $t$ to obtain independent samples when executing the Metropolis-Hastings algorithm and the time-discrete model on the same CPU.}
	\label{fig7}
\end{figure}

This is equivalent to the feedback signal generated in the FPGA of our experimental system in eq.\eqref{eq3}. The time-multiplexed system is therefore equivalent to a numerical integration with a fixed step width of $h=\tau_l$. To demonstrate that this serves as an adequate approximation of the time-continuous model, we show the Kullback-Leibler divergence for the random sparse graphs in Fig.\ref{fig7}a for $\alpha=0.9$ and $\beta=0.1$. Compared to the simulation with a smaller step width in Fig.\ref{fig4}a (where $h=0.1$ was chosen), we find that the sampling accuracy is mostly unaffected by the larger step width in the time-multiplexed model. In Fig.\ref{fig7}b, we show the number of iterations as well as the CPU runtime for obtaining statistically independent samples when simulating eq.\eqref{eq_s24} on a CPU. Compared to Fig.\ref{fig5}b, we find that the overall number of iterations needed to obtain statistically independent samples is slightly reduced by the larger step width. Overall, the sampling accuracy and scaling are well-matched with the simulation of the time-continuous system in Fig.\ref{fig5}b and Fig.\ref{fig4}a. We can therefore conclude that the numerical findings can also be extended to our experimental system as well as to other time-multiplexed Ising machines \cite{INA16a}. 

\subsection*{Runtime analysis of Metropolis-Hastings algorithm and Ising machine simulations on a CPU}

Based on the number of iterations required to obtain statistically independent samples in Fig.\ref{fig5}b we analyze the runtime of both MCMC and Ising machine simulations on a CPU. For the Ising machine simulations, the differential equation \eqref{eq_s2} is integrated using the forward Euler method with a stepwidth of $h=0.1$. MCMC simulations are performed using the Metropolis-Hastings algorithm. This is an iterative algorithm, that updates a random single spin $\sigma_m$ during each iteration according to $\{\sigma\}_\mathrm{new}=\{...,\sigma_{m,\mathrm{new}},...\}=\{...,-\sigma_{m,\mathrm{old}},...\}$. After changing the sign of the randomly selected spin, the energy before $E_\mathrm{old}$ and after the update $E_\mathrm{new}$ are compared. The update is accepted if $E_\mathrm{new}\leq E_\mathrm{old}$. If $E_\mathrm{new}>E_\mathrm{old}$, then the spin update will be accepted with a probability of

\begin{equation}
P=e^{\frac{-(E_\mathrm{new}-E_\mathrm{old})}{T}} \ .
\end{equation}

Otherwise, the spin state will remain unchanged for the next iteration. 

The average runtime of both algorithms is estimated from the average time required for performing one single iteration on a Intel Core i7-7700HQ CPU and then multiplying it with the number of steps in Fig.\ref{fig5}b. The runtime for a single iteration is obtained by repeatedly running both algorithms for 100,000 iterations for all graph sizes from $N=16$ to $N=8192$ and then dividing the averaged runtime by the number of iterations. In general, the time required for performing a single Monte-Carlo step is around 5 times faster with than for performing a single iteration with the Ising machine simulation. Still, we find that reduced number of required iterations results in a significantly faster run time for sampling with the Ising machine simulation.

\subsection*{Mapping of restricted Boltzmann machines to Ising machines}

Restricted Boltzmann machines (RBMs) are Boltzmann machines with a bipartite connection structure. They consist of visible neurons $v_m=\left\{0,1\right\}$ and hidden neurons $h_m=\left\{0,1\right\}$, which are organized into two distinct layers. The layers are fully interconnected with each other while no connections exist within the layers. By stacking several RBMs, a deep belief network (DBN) can be formed, where visible neurons serve as the input and hidden neurons serve as output layers that feed into the next layer. An important step in training RBMs and DBNs is layerwise unsupervised training. During unsupervised training, the logarithmic likelihood of the activation probabilities of the neurons is maximized in each RBM layer in relation to the training data. The biases for the visible neurons $b_{m,\mathrm{vis}}$, hidden neurons $b_{m,\mathrm{hid}}$ and the connection weights $w_{mn}$ are optimized in a gradient descent with the learning rate $\epsilon$ during each training iteration $l$ according to

\begin{equation}
	w_{mn}[l]=w_{mn}[l-1]+\epsilon \left(\left\langle v_m h_n\right\rangle_{\mathrm{data}} - \left\langle v_m h_n\right\rangle_{\mathrm{model}} \right)
\end{equation}
\begin{equation}
	b_{m,\mathrm{vis}}[l]=b_{m,\mathrm{vis}}[l-1]+\epsilon \left(\left\langle v_m \right\rangle_{\mathrm{data}} - \left\langle v_m \right\rangle_{\mathrm{model}} \right) 
\end{equation}
\begin{equation}
	b_{m,\mathrm{hid}}[l]=b_{m,\mathrm{hid}}[l-1]+\epsilon \left(\left\langle h_m \right\rangle_{\mathrm{data}} - \left\langle h_m \right\rangle_{\mathrm{model}} \right) \ \ . \label{eq4}
\end{equation}

$\left\langle ... \right\rangle_{\mathrm{data}}$ and $\left\langle ... \right\rangle_{\mathrm{model}}$ denote the expectation values of the neuron activation probabilities for the RBM with and without training data injected into the input layer respectively. While $\left\langle ... \right\rangle_{\mathrm{data}}$ can be calculated in a straightforward way \cite{ACK85}, the expectation values for the free-running system $\left\langle ... \right\rangle_{\mathrm{model}}$ have to be approximated through Boltzmann sampling. In general, this sampling is considered NP-hard and computationally expensive to do on digital computers. Instead, less accurate estimation methods that can only be applied to RBMs, such as contrastive divergence, are often employed \cite{CAR05}. 

The energy of RBMs is calculated similar to the Ising Hamiltonian according to

\begin{equation}
	E_\mathrm{RBM}=-\sum_{mn}{w_{mn}v_m h_n}-\sum_m{b_{m,\mathrm{vis}}v_m}-\sum_m{b_{n,\mathrm{hid}}h_n} \label{eq_s3}
\end{equation}

To map the binary neurons $v_m=\left\{0,1\right\}$ and $h_m=\left\{0,1\right\}$ to an Ising spin model $\sigma_m=\left\{-1,1\right\}$, we employ the linear relations $v_m=(\sigma_{m,\mathrm{vis}}+1)/2$ and $h_n=(\sigma_{n,\mathrm{hid}}+1)/2$. Inserting these relations into equation \eqref{eq_s3}, we obtain the Ising Hamiltonian corresponding to the restricted Boltzmann machine

mathrm\begin{equation}
	E_{\mathrm{RBM,Ising}}=-\frac{1}{4}\sum_{mn}{w_{mn}\sigma_{m,\mathrm{vis}}\sigma_{n,\mathrm{hid}}}-\frac{1}{4}\sum_{mn}w_{mn}\sigma_{m,\mathrm{vis}}-\frac{1}{4}\sum_{mn}w_{mn}\sigma_{n,\mathrm{hid}}-\frac{1}{2}\sum_m{b_{m,\mathrm{vis}}\sigma_{m,\mathrm{vis}}}-\frac{1}{2}\sum_m{b_{n,\mathrm{hid}}\sigma_{n,\mathrm{hid}}} \ \ . \label{eq_s4}
\end{equation}

Here, we exploit the invariance of the Ising Hamiltonian under constant energy shifts to remove the constant energy terms from equation \eqref{eq_s4}. Comparing equation \eqref{eq_s4} to the Ising Hamiltonian, we find that the first term corresponds to the coupling term, where $w_{mn}$ is equivalent to $J_{mn}$. The last four terms then are equivalent to the bias term in equation \eqref{eq2}. The $N$ hidden and $M$ visible neurons are expressed as a combined spin state $\{\sigma\}=\{\sigma_{1,\mathrm{vis}},...,\sigma_{N,\mathrm{vis}},\sigma_{1,\mathrm{hid}},...,\sigma_{M,\mathrm{hid}}\}$.

\subsection*{RBM training algorithm}

The RBM and its training are implemented with the scikit-learn machine learning library for Python 3. The training data is generated from the digits dataset \cite{ALI97}. To increase the difficulty and to increase the size of the training dataset, each image is additionally shifted by one pixel into all four spatial directions. This results in a five times larger dataset containing 8985 images. The dataset is split into 7188 images for training and 1797 images for testing. For each training iteration, the RBM is trained in an unsupervised way. After the unsupervised training step, a logistic regression layer is trained on the output of the RBM in a supervised way. For this, the training dataset is applied as input to the visible RBM layer. The logistic regression layer uses the newton-cg solver for optimizing its hyperparameters with an inverse regularization strength of 6000. After the supervised training, the prediction accuracy is then measured by using the test dataset. 

\subsection*{Implementation of noise-induced Boltzmann sampling with general gain-dissipative systems}

\begin{figure}[htbp]
	\centering
		\includegraphics[width=.99\textwidth]{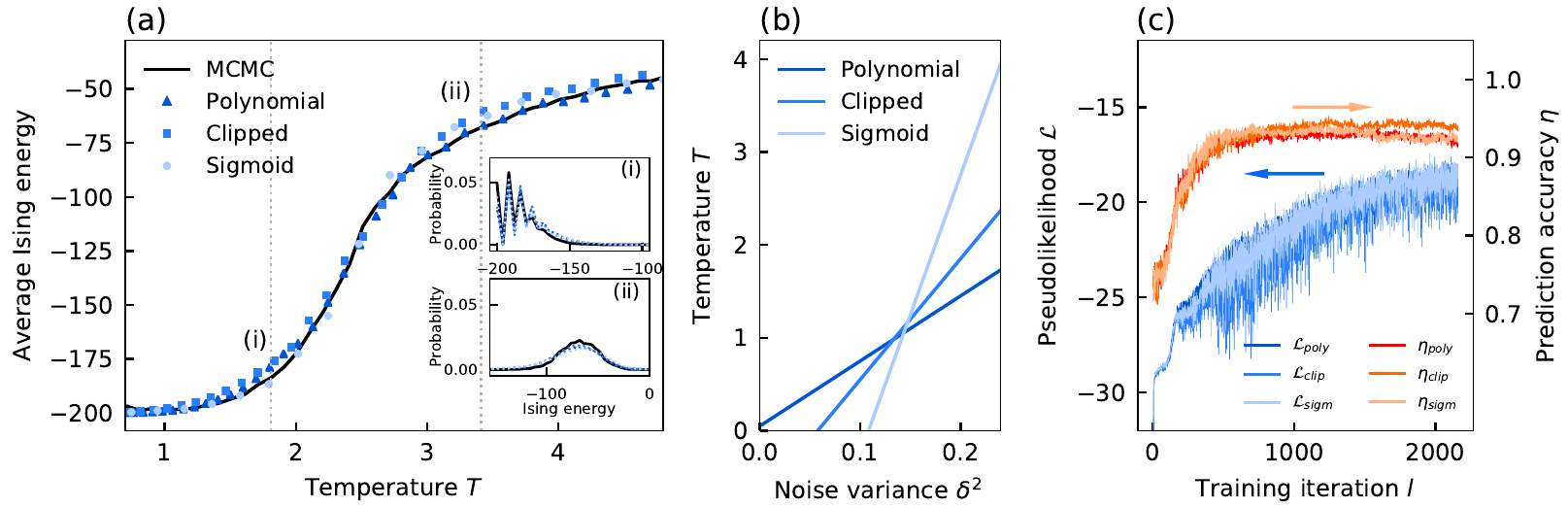}
	\caption{\textbf{Simulated Boltzmann sampling performance of different analog Ising machine models} (a) Average energy at different temperatures for the 2D Ising model. Samples are obtained with the Metropolis-Hastings algorithm (MCMC, solid line) and with Ising machine simulations for different gain-dissipative systems (triangle: polynomial, square: clipped, circle: sigmoid). Insets: Energy distributions for the four different systems at $T=1.8$ (i) and $T=3.4$ (ii). (b) Relation between noise variance $\delta^2$ and the temperature for the 2D Ising model for the different gain-dissipative systems. (c) Unsupervised training for the digit recognition task in Fig.4 using simulations of different gain-dissipative systems.}
	\label{fig8}
\end{figure}

To show the universality of noise-induced sampling, we consider its implementation on various types of analog Ising machines. In general, analog Ising machines can be realized with a number of gain-dissipative systems, ranging from optical parametric oscillators to electronic oscillators. To model the temporal evolution of these systems, the equations of motion can be unified into a generalized model \cite{BOE21}. Different systems are then expressed through their nonlinear transfer function, which comprise a variety of function classes. In addition to the clipped nonlinearity used for modeling opto-electronic system (equation \eqref{eq_s2}), we consider gain-dissipative systems based on polynomial and sigmoid nonlinearities. Gain-dissipative systems based on polynomial functions are commonly used in the modeling of analog Ising machines based on optical parametric oscillators, polariton condensates or ring resonators. Their equation of motion is given by

\begin{equation}
	\frac{dx_m}{dt}=(\alpha-1)x_m-x_m^3+\beta\left(\sum_{n}J_{mn}x_n+x_\mathrm{sat} b_m\right)+\delta\zeta(t) \ \ . \label{eq_s5}
\end{equation}

Gain-dissipative systems based on sigmoid functions are common in the modeling of neural networks but have recently also been identified as a good choice for building analog Ising machines \cite{BOE21}. Their equation of motion is given by

\begin{equation}
	\frac{dx_m}{dt}=-x_m+\tanh(\alpha x_m+\beta\left(\sum_{n}J_{mn}x_n+x_\mathrm{sat} b_m\right)+\delta\zeta(t)) \ \ . \label{eq_s6}
\end{equation}

To test the implementation of noise-induced sampling with the different systems, we perform simulations of the equations of motion using the forward Euler method. First, we perform noise induced sampling of a 2D Ising model with $N=100$ spins at different temperatures. In the 2D Ising model, spins are arranged in a grid structure and coupled to their four nearest neighbors. It is a common benchmark for statistical sampling, as it possesses a second order phase transition and therefore exhibits critical phenomena, such as critical slowing down. We perform noise-induced sampling by simulating equation \eqref{eq_s2}, equation \eqref{eq_s5} and equation \eqref{eq_s6} for different noise variances $\delta$. The simulations are performed with an integration step width of $h=0.1$ for 90000 iterations at $\alpha=0.8$ and $\beta=0.1$. In Fig.\ref{fig8}, we show the average Ising energy obtained from the different systems in comparison to MCMC-based sampling using the Metropolis-Hastings algorithm. Over the entire temperature range and specifically at the phase transition point $T_\mathrm{crit}\approx 2.27$, we find a good agreement between software-based and Ising machine-based sampling. In the insets in Fig.\ref{fig8}, we show exemplary comparisons of the sampled energy distribution of the different models. Compared to MCMC-based sampling, the different analog Ising machine models provide good approximations of the distribution and hence demonstrate that noise-induced sampling can provide accurate Boltzmann sampling independent of the specific type of analog Ising machine. In Fig.\ref{fig8}b, we show the temperature as function of the noise variance in the 2D Ising model for the different systems. As for the 4-spin system in Fig.2, we again observe a linear relation between the noise variance and the temperature of the Ising model. 

Finally, we consider the different systems for the unsupervised training task in Fig.4. Here, we perform sampling through simulations of equation \eqref{eq_s2}, equation \eqref{eq_s5} and equation \eqref{eq_s6} with a step width of $h=1$ at $\alpha=0.9$ and $\beta=0.1$ for 1000 steps. The machine learning model is identical to the one in Fig.4 and is trained with a learning rate of $\epsilon=0.1$. For each model, the noise level is adjusted to $\delta_\mathrm{poly}=0.13$, $\delta_\mathrm{clip}=0.12$ and $\delta_\mathrm{sigm}=0.19$. In Fig.\ref{fig8}, we track the pseudolikelihood and the prediction accuracy as a function of the training iteration. When comparing the different models, we note that their performance is almost identical with slight differences in the maximum predication accuracy ($\mathrm{max}(\eta_\mathrm{poly})=0.938$, $\mathrm{max}(\eta_\mathrm{clip})=0.953$ and $\mathrm{max}(\eta_\mathrm{sigm})=0.944$). Overall, we find that unsupervised training can be performed on a variety of analog Ising machines through noise-induced sampling.

\section*{Data availability}
The authors declare that all data needed to evaluate the conclusions of the paper are present in the paper. The sparse random graphs files have been deposited in the Open Science Framework under 'Böhm, F. Noise-injected analog Ising machines enable ultrafast statistical sampling and machine learning. (2022). doi:10.17605/OSF.IO/347XT'. Additional data are available from the corresponding authors upon reasonable request.

\section*{Code availability}
The modified scikit-learn library files for training RBMs with Ising machines have been deposited in the Open Science Framework under 'Böhm, F. Noise-injected analog Ising machines enable ultrafast statistical sampling and machine learning. (2022). doi:10.17605/OSF.IO/347XT'.

\section*{Acknowledgments}
We acknowledge financial support from the Research Foundation Flanders (FWO) under the grants G028618N, G029519N, and G006020N as well as the Hercules Foundation and the Research Council of the Vrije Universiteit Brussel (F.B., G.V., G.V.d.S.). Additional funding was provided by the EOS project "Photonic Ising Machines". This project (EOS number 40007536) has received funding from the FWO and F.R.S.-FNRS under the Excellence of Science (EOS) programme (F.B., G.V., G.V.d.S.).

\section*{Author contributions}
F.B. designed and performed the experiments and analyzed the data. F.B. and D.A.U. performed and analyzed the simulations. F.B., D.A.U., G.V. and G.V.d.S. discussed the results and wrote the paper.

\section*{Competing interests}
The authors declare no competing interests.

\end{document}